\documentclass[twocolumn,preprintnumbers,elsart]{revtex4}
\usepackage{amssymb}
\usepackage{amsmath}
\usepackage{graphicx}
\usepackage{dcolumn}
\usepackage[center]{subfigure}
\usepackage{array}
\usepackage{booktabs}
\usepackage{multirow}
\usepackage{color}

\begin{document}

\title{Discrete solitons in zigzag waveguide arrays with different types of
linear mixing between nearest-neighbor and next-nearest-neighbor couplings}
\author{Jinzhou Hu$^{1,\S }$, Shulan Li$^{1,\S }$, Zhaopin Chen$^{2}$,
Jiantao L\"{u}$^{1}$, Bin Liu$^{1}$}
\email{binliu@fosu.edu.cn}
\author{Yongyao Li$^{1,2}$}
\affiliation{$^{1}$School of Physics and Optoelectronic Engineering, Foshan University,
Foshan 528000, China}
\affiliation{$^{2}$ Department of Physical Electronics, School of Electrical Engineering,
Faculty of Engineering, Tel Aviv University, Tel Aviv 69978, Israel}
\affiliation{$^{\S }$ These authors contributed equally to this work.}

\begin{abstract}
We study discrete solitons in zigzag discrete waveguide arrays with
different types of linear mixing between nearest-neighbor and
next-nearest-neighbor couplings. The waveguide array is constructed from two
layers of one-dimensional (1D) waveguide arrays arranged in zigzag form. If
we alternately label the number of waveguides between the two layers, the
cross-layer couplings (which couple one waveguide in one layer with two
adjacent waveguides in the other layer) construct the nearest-neighbor
couplings, while the couplings that couple this waveguide with the two
nearest-neighbor waveguides in the same layer, i.e., self-layer couplings,
contribute the next-nearest-neighbor couplings. Two families of discrete
solitons are found when these couplings feature different types of linear
mixing. As the total power is increased, a phase transition of the second
kind occurs for discrete solitons in one type of setting, which is formed
when the nearest-neighbor coupling and next-nearest-neighbor coupling
feature positive and negative linear mixing, respectively. The mobilities
and collisions of these two families of solitons are discussed
systematically throughout the paper, revealing that the width of the soliton
plays an important role in its motion. Moreover, the phase transition
strongly influences the motions and collisions of the solitons.
\end{abstract}

\maketitle

\section{Introduction}

The transmission of light fields in discrete systems manifests abundant
functional phenomena \cite{Kivshar2012}. Discrete waveguide arrays represent
the most essential and fundamental element for discrete optics \cite{DNC2003}%
. Diffraction management \cite{HSE2000}, including anomalous refraction,
negative refraction, Anderson location \cite{Lahini2008}, and Bloch
oscillation \cite{Pertsch21999}, have been reported experimentally in
discrete waveguide arrays. Another important phenomenon in nonlinear optical
waveguide arrays is produced by the self-trapping of light, namely, discrete
solitons, which are created through a balance between discrete diffraction
effects and nonlinearity \cite%
{Lederer2008,Eisenberg1998,RMorandotti,Xiangyu2014,Zhiqiang,DCQ1,DCQ2,XUTAO1,XUTAO2,YFW,YZY,oe14_12347,PRA78_011804}%
. Discrete solitons in nonlinear waveguide array networks can provide an
ideal platform for all-optical data processing applications and can achieve
intelligent functional operations such as routing, blocking, logic
functions, and time-gating \cite{DNC2001}. Hence, discrete waveguide arrays
are important devices for all-optical switching networks, similar to
semiconductor devices in electronic circuits, and the discrete solitons in
different functional discrete waveguide systems constitute an active topic
in optics.

Generally, the tunneling of the light field among different waveguides
originates from the evanescent coupling between adjacent (i.e.,
nearest-neighbor) waveguides, whereas the evanescent coupling between
next-nearest-neighbor waveguides can be neglected. Hence, the propagation of
a light field in this kind of waveguide array is always described by the
discrete nonlinear Schr\"{o}dinger equation (DNSE) with only first-order
discrete diffraction. However, it was reported that next-nearest-neighbor
coupling was excited by the hopping of the light field in zigzag waveguide
arrays \cite{NKE2002,PGK2003,Szameit,CChong2011,Dovgiy2015}. Long-range
coupling is introduced because a zigzag arrangement enhances the hopping
rate of the light field to the next-nearest-neighbor site, and the
propagation of a light field in zigzag waveguide arrays can be described by
the DNSE with second-order discrete diffraction. Recently, a discrete
waveguide array with an exponentially long range coupled effect was reported
to be realized with the special design of the waveguide system \cite%
{Zhijie2014,Zhijie2015}. Because the coupling between the waveguides
indicates a transition of energy, it always features negative linear mixing
in the field between two connecting waveguides. In comparison, the coupling
featuring positive linear mixing between two connecting waveguides has
rarely been discussed. Recently, the synthetic (artificial) gauge field or
complex coupling, which makes the coupling constant a complex number, was
introduced into optical systems \cite%
{Moti,Hongxing,Shanhui2012,Golshani,Ardakani,Nezhad2017}, allowing the
coupling between two connecting waveguides to be tuned to feature either
negative or positive linear mixing.

\begin{figure}[h]
{\includegraphics[width=1.0\columnwidth]{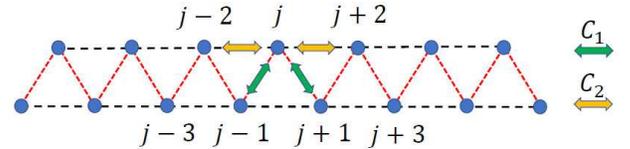}}
\caption{Sketch of the zigzag waveguide arrays. Here, $C_{1}$ and $C_{2}$
denote the nearest-neighbor and next-nearest-neighbor couplings,
respectively. In this model, these two couplings satisfy $C_{1}\cdot C_{2}<0$%
.}
\label{sketch}
\end{figure}

In this paper, we consider the zigzag waveguide system shown in Fig. \ref%
{sketch}. This waveguide array is formed by two layers of one-dimensional
(1D) waveguide arrays. The waveguide in one layer constructs an equilateral
triangle with the two adjacent waveguides of the other layers. If we
alternately label the number of waveguides between two layers of the
waveguides, which are ordered in the same sequence as in Ref. \cite{NKE2002}%
, the cross-core couplings, which couple one waveguide in one layer with the
two adjacent waveguides in the neighboring layer, construct the
nearest-neighbor coupling. In contrast, the self-core couplings, which
couple the waveguide in one layer with the two nearest-neighbor waveguides
in the same layers, contribute next-nearest-neighbor couplings. By
introducing the synthetic gauge field into the coupling coefficient, we
assume that these two types of couplings, i.e., cross-coupling and
self-coupling, feature opposite types of linear mixing (i.e., negative and
positive linear mixing) between the waveguides that are coupled. Hence, the
aim of this paper is to study the characteristics, mobilities, and
interactions of discrete solitons modulated by the opposite linear mixing
between these two types of coupling. The rest of the paper is structured as
follows: the model is described in Section II, the analysis and numerical
results of the stationary solution and the dynamics of the discrete solitons
are presented in Sections III and IV, respectively, and the conclusions are
drawn in Section V.

\section{The model}

The settings of our model are illustrated in Fig. \ref{sketch}. The zigzag
waveguide is formed by two layers of $1$D waveguide arrays. The waveguide in
one layer constructs an equilateral triangle with the two adjacent
waveguides of the other layers. The sequence of the waveguide number is
defined as the sketch, which follows the same definition as in Ref. \cite%
{NKE2002}. Hence, the nearest-neighbor couplings for a waveguide in one
layer, namely, $C_{1}$, are contributed from the two adjacent waveguides in
the neighboring layer, while the next-nearest-neighbor couplings, namely, $%
C_{2}$, are contributed from the closest neighbor waveguides in the same
layer. We assume that these waveguides feature self-focusing Kerr
nonlinearity. The Hamiltonian of this system, in scaled form, is defined as
\cite{NKE2002}
\begin{equation}
H={\frac{1}{2}}\sum_{n}\left(
C_{1}|u_{n}-u_{n-1}|^{2}+C_{2}|u_{n}-u_{n-2}|^{2}-|u_{n}|^{4}\right) ,
\label{Ham}
\end{equation}%
where $u_{n}$ is the dimensionless amplitude of the field in the $n$-th
waveguide.

The propagation of the light field in the current system is governed by the
DNSE, which can be obtained by $idu_{n}/dz=\partial H/\partial u_{n}^{\ast }$
as
\begin{eqnarray}
&&i{\frac{d}{dz}}u_{n}=-{\frac{C_{1}}{2}}\left(
u_{n+1}+u_{n-1}-2u_{n}\right)   \notag \\
&&-{\frac{C_{2}}{2}}\left( u_{n+2}+u_{n-2}-2u_{n}\right) -|u_{n}|^{2}u_{n},
\label{DNLS}
\end{eqnarray}

The total power of the field is
\begin{equation}
P=\sum_{n} |u_{n}|^{2}.  \label{P}
\end{equation}

The stationary soliton solutions of Eq. (\ref{DNLS}) can be written as
\begin{equation}
u_{n}(z)=\phi _{n}e^{i\beta z},  \label{uU}
\end{equation}%
where $\phi _{n}$ is the dimensionless amplitude of the soliton and $\beta $
is the propagation constant. The stability of the localized stationary modes
is investigated numerically by means of computing the eigenvalues for small
perturbations and is verified by proforming direct simulations. The
perturbed solution is taken as
\begin{equation}
u_{n}=e^{i\beta z}(\phi _{n}+w_{n}e^{i\lambda z}+v_{n}^{\ast }e^{-i\lambda
^{\ast }z}),
\end{equation}%
where the asterisk denotes the complex conjugate. Following the substitution
of this expression into Eq. (\ref{DNLS}), linearization leads to the
eigenvalue problem for $\lambda $ and the eigenmodes $\left(
w_{n},v_{n}\right) $:
\begin{equation}
\left(
\begin{array}{cc}
\mathbf{C}+\beta -2|\phi _{n}|^{2} & \phi _{n}^{2} \\
-\phi _{n}^{\ast 2} & -\mathbf{C}-\beta +2|\phi _{n}|^{2}%
\end{array}%
\right) \left(
\begin{array}{c}
w_{n} \\
v_{n}%
\end{array}%
\right) =\lambda \left(
\begin{array}{c}
w_{n} \\
v_{n}%
\end{array}%
\right) .  \label{eigen}
\end{equation}%
where $\mathbf{C}$ is a matrix that defines the total effect of the linear
coupling. The elements of $\mathbf{C}$ are defined as
\begin{eqnarray}
&&C_{ij}=-{\frac{1}{2}}C_{1}\left( \delta _{i,j-1}+\delta _{i,j+1}-2\delta
_{i,j}\right)   \notag \\
&&-{\frac{1}{2}}C_{2}\left( \delta _{i,j-2}+\delta _{i,j+2}-2\delta
_{i,j}\right) ,
\end{eqnarray}%
where $\delta _{i,j}$ is the Kronecker symbol. The solution $\phi _{n}$ is
stable if the spectrum of the eigenvalues $\lambda $ is real. If we select
the waveguide arrays fabricated on AlGaAs, and the real nonlinear parameter
and real coupling coefficient are $3.6$ m$^{-1}$W$^{-1}$ and 0.82 mm$^{-1}$,
respectively \cite{Eisenberg1998,Aitchison1997}, the unit of the scaled total power, i.e., $P=1$, is 4.4
kW if the coupling coefficient is normalized to $1$ by rescaling the
nonlinear strength. The coupling coefficient of $C=1$ corresponds to a
distance of $\sim 5$ microns between two coupled waveguides \cite{Eisenberg1998}.

Here, positive and negative values of the coupling constant $C_{i}$\ (where $%
i=1$\ or $2$) indicate that these two linear couplings feature an opposite
linear mixing. In the current system, because the two types of coupling are
opposite to each other, the coupling constants $C_{1,2}$ satisfy $C_{1}\cdot
C_{2}<0$. In the next section, we carry out analytical and numerical studies
of the stationary discrete solitons in the present system.

\section{Characteristics of the soliton}

\subsection{Analysis}

\begin{figure}[h]
{\includegraphics[width=0.8\columnwidth]{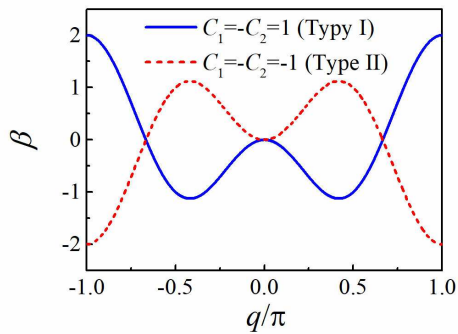}}
\caption{Dispersion curves for the cases of $(C_{1},C_{2})=(1,-1)$ (type I,
blue solid line) and $(-1,1)$ (type II, red dashed line).}
\label{disperfig}
\end{figure}

For convenience, we select $\left\vert C_{1}\right\vert =$ $\left\vert
C_{2}\right\vert =1$ in Eq. (\ref{DNLS}). Note again that $C_{1}$ represents
the coupling between two adjacent layers and $C_{2}$ represents the
couplings within the same layer, which can be characterized as
cross-coupling and self-coupling, respectively. Linearizing Eq. (\ref{DNLS})
by the plane wave $u_{n}=A\exp \left( i\beta z+iqn\right) $ (where $A$ is
the real amplitude of the plane wave), one can obtain the following
dispersion relation for the current system:
\begin{equation*}
\beta =C_{1}\cos q+C_{2}\cos 2q-(C_{1}+C_{2}).
\end{equation*}%
Because $C_{1,2}$ manifests opposite types of linear mixing, two types of
systems are created when $(C_{1},C_{2})=(1,-1)$ and $(-1,1)$, respectively.
The dispersion curves in the first Brillouin zone (i.e., $-\pi \leq q\leq
\pi $) for these two cases are displayed in Fig. \ref{disperfig}.

The dispersion curves indicate that these two types of systems feature
opposite dispersion relations at the center (i.e., $q=0$) and the edge
(i.e., $q=\pi $) of the Brillouin zone. At the center of the Brillouin zone,
the effective diffraction of the type I array is \textquotedblleft normal"
(i.e., $\beta ^{\prime \prime }(q)|_{q=0}>0$), while that of the type II
array is \textquotedblleft anomalous" (i.e., $\beta ^{\prime \prime
}(q)|_{q=0}<0$). At the edge of the Brillouin zone, the types of effective
diffraction are reversed: the type I array becomes \textquotedblleft
anomalous" (i.e., $\beta ^{\prime \prime }(q)|_{q=\pi }<0$), and the type II
array becomes \textquotedblleft normal" (i.e., $\beta ^{\prime \prime
}(q)|_{q=\pi }>0$). Even though the stationary solutions in Eq. (\ref{DNLS})
are solved only numerically, we can still give some rough analysis of the solitons at the
center and edge of the Brillouin zone for the two types of current systems.
We assume that the discrete soliton can be written as
\begin{equation*}
u_{n}=A_{n}e^{i\beta z+iqn},
\end{equation*}%
where $A_{n}$ is the real amplitude and $q$ is equal to $0$ or $\pi $. At
the center of the Brillouin zone (i.e., $q=0$), the discrete solitons in the
type I and type II arrays are described by the following two stationary
equations, respectively:
\begin{eqnarray}
-\beta A_{n} &=&-{\frac{1}{2}}\left( A_{n+1}+A_{n-1}\right) +{\frac{1}{2}}%
\left( A_{n+2}+A_{n-2}\right)   \notag \\
&&-A_{n}^{3},  \label{q0I} \\
-\beta A_{n} &=&{\frac{1}{2}}\left( A_{n+1}+A_{n-1}\right) -{\frac{1}{2}}%
\left( A_{n+2}+A_{n-2}\right)   \notag \\
&&-A_{n}^{3}.  \label{q0II}
\end{eqnarray}%
Similarly, at the edge of the Brillouin zone (i.e., $q=\pi $), the discrete
solitons in these two types of arrays are described by
\begin{eqnarray}
&&-\beta A_{n}={\frac{1}{2}}\left( A_{n+1}+A_{n-1}\right) +{\frac{1}{2}}%
\left( A_{n+2}+A_{n-2}\right)   \notag \\
&&\text{ \ \ \ \ \ \ \ \ }-A_{n}^{3},  \label{qpiI} \\
&&-\beta A_{n}=-{\frac{1}{2}}\left( A_{n+1}+A_{n-1}\right) -{\frac{1}{2}}%
\left( A_{n+2}+A_{n-2}\right)   \notag \\
&&\text{ \ \ \ \ \ \ \ \ }-A_{n}^{3}.  \label{qpiII}
\end{eqnarray}
Eqs. (\ref{q0II}) and (\ref{qpiII}) can be equivalent to discrete solitons
at the edge ($q=\pi $) and center ($q=0$) of the Brillouin zone of zigzag
waveguide arrays with $C_{1}\cdot C_{2}>0$ \cite{NKE2002}. Bright solitons
in these two cases are staggered and unstaggered, respectively. Eq. (\ref%
{qpiI}) can support a bright staggered soliton solution if we apply a
transformation of $\tilde{A}_{n}=-A_{n}$ \cite{NKE2002}, which can also be
indirectly demonstrated by Fig. \ref{kickpi}(c) in Section IV. Moreover, Eq.
(\ref{qpiI}) can support dark (or gray) discrete solitons after such a
transformation \cite{NKE2002}. To the best of our knowledge, Eq. (\ref{q0I})
is never considered for discrete solitons. In the following subsection, we
carry out numerical simulations and consider the discrete soliton at the
center of the Brillouin zone for both types of systems with $%
(C_{1},C_{2})=(1,-1)$ and $(-1,1)$.

\subsection{Numerical results}

Discrete solitons at the center of the Brillouin zone are ground-state
solutions, which can be numerically solved by the imaginary-time method
(ITM) (a standard algorithm for searching for the ground-state solutions of
the nonlinear Schr\"{o}dinger equation and Gross-Pitaevskii equation) \cite%
{Chiofalo,Jianke}. Two families of discrete solitons can be created through
the current system with two types of opposite linear mixing between the
nearest-neighbor and next-nearest-neighbor couplings. The characteristics of
these two families of discrete solitons for these two types of systems are
listed in Table I.

\begin{table}[h]
\caption{Discrete solitons in the two types of systems }
{\
\begin{tabular}{ccc}
\hline\hline
\specialrule{0em}{1pt}{1pt} Type & $(C_{1},C_{2})$ & Description \\ \hline
\specialrule{0em}{1pt}{1pt} Type I & $(1,-1)$ & Cross coupling is negative;
\\
(Multipeak) &  & self coupling is positive. \\
\specialrule{0em}{1pt}{1pt} Type II & $(-1,1)$ & Cross coupling is positive;
\\
(Staggered) &  & self coupling is negative. \\ \hline
\end{tabular}%
} 
\label{tableI}
\end{table}

\begin{figure}[h]
{\includegraphics[width=1\columnwidth]{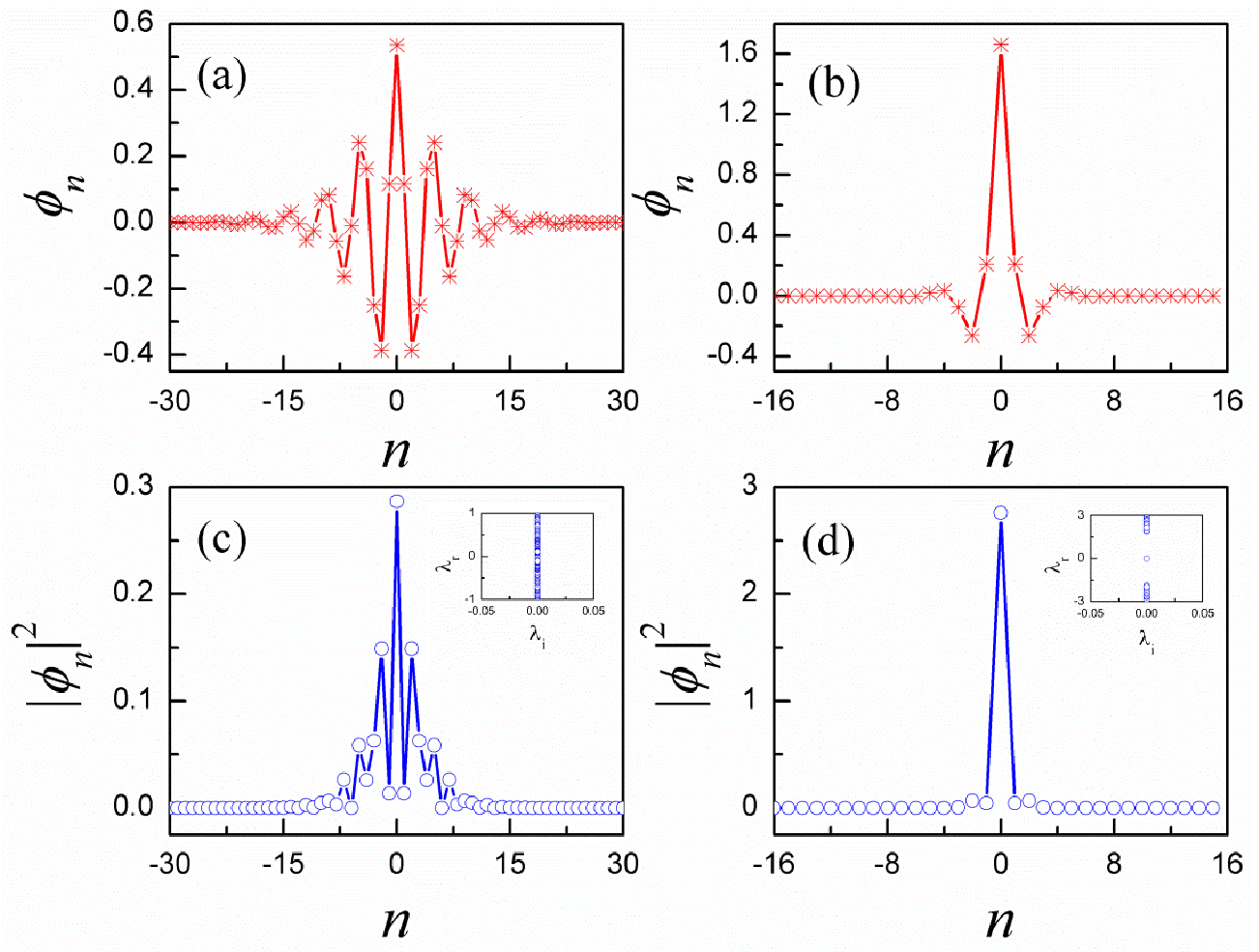}}
\caption{Typical example of the discrete soliton in the current system with $%
(C_{1},C_{2})=(1,-1)$ (type I system). (a,b) Amplitudes of the soliton with $%
P=1.0$ and $3.0$, respectively. (c,d) Intensities of the soliton in panels
(a,b), respectively. The insets are the spectra of $\protect\lambda $, which
demonstrate the stabilities of these solitons. }
\label{ExpTypeI}
\end{figure}

\begin{figure}[h]
{\includegraphics[width=1.0\columnwidth]{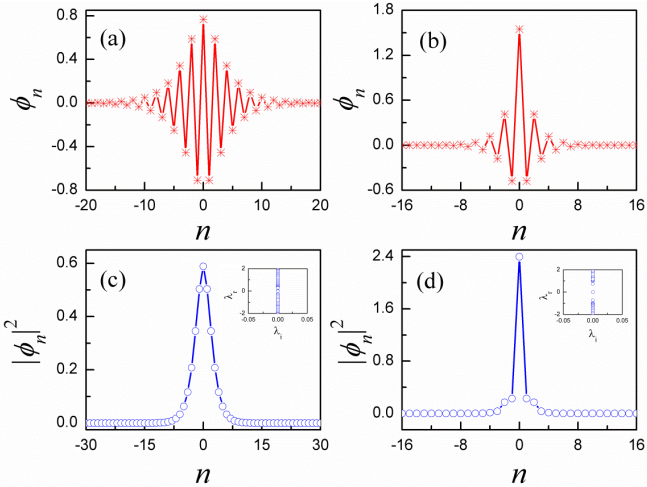}}
\caption{Typical example of the discrete soliton in the current system with $%
(C_{1},C_{2})=(-1,1)$ (type II system). (a,b) Amplitudes of the soliton with
$P=3.2$ and $3.3$, respectively. (c,d)Intensities of the soliton in panels
(a,b), respectively. The insets are the spectra of $\protect\lambda $, which
demonstrate the stabilities of these solitons. }
\label{ExpTypeII}
\end{figure}

Typical examples of stable solitons for type I and II systems are displayed
in Figs. \ref{ExpTypeI} and \ref{ExpTypeII}, respectively. These solitons
exhibit different characteristics in terms of their amplitudes and
intensities. The intensity of the soliton in the type I system features a
multipeak structure, while that of the soliton in the type II system
features a single-peak structure. The amplitudes of the soliton in the type
I system show a $\pi $ phase shift between two adjacent peaks, while the
soliton in the type II system display a staggered structure at each lattice
site (which is in accordance with the analysis in the above subsection).
Without considering the next-nearest-neighbor coupling, a $\pi $ phase shift
for peaks or staggered discrete solitons always exists for lattice gap
solitons or discrete solitons with self-defocusing Kerr nonlinearity. Here,
we observe them in the self-focusing Kerr nonlinearity via the current
system. Note that the staggered solitons in the current system are different
from their counterparts in Ref. \cite{NKE2002}, which are created at the
edge of the Brillioun zone (i.e., $q=\pi $); in contrast, the current
staggered solitons are created at the center of the Brillioun zone (i.e., $%
q=0$), which results in a zigzag waveguide array with opposite linear mixing
between the two types of coupling. If we continue to increase the power, the
energy of the field shrinks to the scale of a single waveguide, and the two
families of solitons become similar.

\begin{figure}[h]
{\includegraphics[width=1.0\columnwidth]{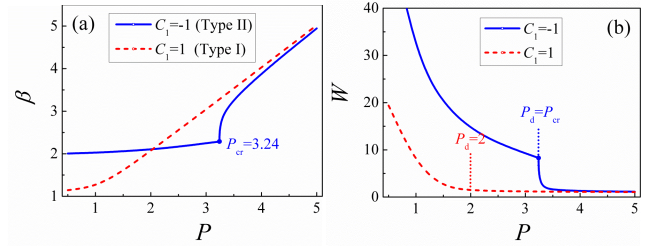}}
\caption{(a) Propagation constant $\protect\beta $ of the soliton versus $P$%
. The solid blue dot ($P=P_{\mathrm{cr}}$) on the curve $\protect\beta (P)$
of the type II soliton indicates the phase transition point. (b) Width of
the soliton versus $P$. The two vertical dotted lines with $P=P_{\mathrm{d}}$
are the borders between the mobile and immobile solitons for the two types
of discrete solitons. On the left side of the dotted line ($P<P_{\mathrm{d}}$%
), the soliton is mobile, while on the right side of the dotted line ($P>P_{%
\mathrm{d}}$), the soliton becomes immobile. Specifically, for the type II
soliton, $P_{\mathrm{d}}=P_{\mathrm{cr}}$.}
\label{character}
\end{figure}

To study the characteristics of the soliton more clearly, we plot the
propagation constant $\beta $ and the effective width as functions of $P$
for the two families of solitons in Fig. \ref{character}, where the
effective width of the soliton is defined as
\begin{equation*}
W={\frac{\left( \sum_{n}|u_{n}|^{2}\right) ^{2}}{\sum_{n}|u_{n}|^{4}}}.
\end{equation*}

The curves of $\beta (P)$ in Fig. \ref{character}(a) for the two families of
solitons show that they satisfy the Vakhitov-Kolokolov (VK) criterion, i.e.,
$d\beta /dP>0$, a necessary stability condition for solitons in
self-focusing media. The curve of $\beta (P)$ for the soliton in the type II
system reveals that a phase transition for this kind of soliton occurs as $P$
is increases, and the phase transition point is located at $P_{\mathrm{cr}%
}=3.24$. The behavior of the curve near the phase transition point indicates
that this phase transition is of the second kind. Unlike the curve of the
soliton in the type II system, the curve of the soliton in the type I system
is smooth, which indicates that a phase transition is not detected as $P$ is
increases. Fig. \ref{ExpTypeII} displays a typical example of the soliton in
the type II system before and after the phase transition. Before the phase
transition (i.e., $P<P_{\mathrm{cr}}$), the intensity profile of the soliton
has a Gaussian shape [see Fig. \ref{ExpTypeII}(c)], which are similar to the
solitons in a continuous system; the solitons in this case can be regarded
as quasi-continuous type. After the phase transition (i.e., $P>P_{\mathrm{cr}%
}$), the intensity profile of the soliton is different from that of the
soliton before the phase transition. The majority of the power of the
soliton shrunks to the central site, which gives rise to a substrate around
the central site [see Fig. \ref{ExpTypeII}(d)]. Then, the curve of $\beta (P)
$ of the soliton in the type II system approaches its counterpart in the
type I system after the phase transition.

The curves of $W(P)$ in Fig. \ref{character}(b) for the two families of
solitons show that $W(P)$ decreases as $P$ increases, which can be naturally
understood by the feature of the solitons in self-focusing Kerr media. A
phase transition is also evident from the curve of $W(P)$ for the soliton in
the type II system, which demonstrates that the width of the soliton
decreases steeply after the phase transition (i.e., $P>P_{\mathrm{cr}}$).

\section{Dynamics of the discrete solitons}

\subsection{Mobility of the solitons}

\begin{figure}[h]
{\includegraphics[width=1.0\columnwidth]{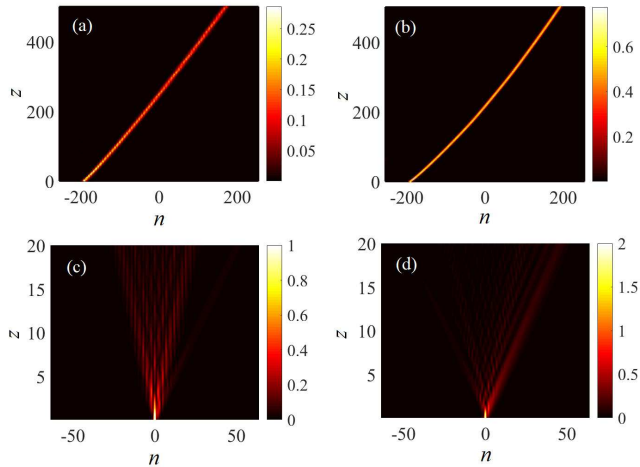}}
\caption{(a) Motion of a discrete soliton in the type I system with $P=1$.
(b) Motion a discrete soliton in the type II system with $P=3.2$. The
strengths of the kicks in panels (a,b) are $\protect\eta =0.1\protect\pi $.
(c) The discrete soliton in the type I system with $P=2.2$ is destroyed by
the strong kick. (d) The discrete soliton in the type II system with $P=3.3$
is destroyed by the strong kick. The strengths of the kick in panels (c,d)
are $\protect\eta =0.3\protect\pi $.}
\label{moving}
\end{figure}

\begin{figure}[h]
{\includegraphics[width=1.0\columnwidth]{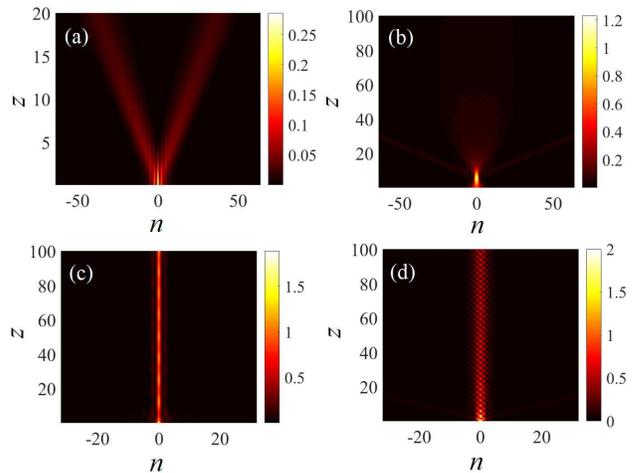}}
\caption{(a,b) Direct simulations of the solitons in the type I and type II
systems with $P=1$ and $P=3.2$, respectively, which are all smaller than $P_{%
\mathrm{d}}$, kicked by $\protect\eta =\protect\pi $. (c,d) Direct
simulations of the soliton in the type I and type II systems with $P=2.2$
and $P=3.3$, respectively, which are all larger than $P_{\mathrm{d}}$,
kicked by $\protect\eta =\protect\pi $.}
\label{kickpi}
\end{figure}

Studying the motion of discrete solitons in a discrete system is a
nontrivial issue for understanding the dynamic of the discrete solitons.
However, how to move a soliton in a discrete lattice while conserving its
shape remains unclear. Many papers have discussed this issue; to date, it
has been found that the width of a discrete soliton strongly influences its
mobility \cite{YVK2009,jinhong,Zhenjing}. Specifically, if a discrete
soliton has a sufficiently broad width (and thus can be termed a
quasi-continuous object), the soliton can be more easily mobilized and can
be partially explained by the Ablowitz-Ladik model \cite{AL}.

In this paper, the motion of the discrete soliton in the current system is
studied by exerting a kick on the stationary solution as
\begin{equation*}
u_{n,z=0}=\phi _{n}e^{i\eta n},
\end{equation*}%
where $\eta $ is the strength of the kick, which can be realized by
imprinting a phase tilt on each waveguide (for example, using a spatial
light modulator for the waveguide). The numerical algorithm that is employed
to study the dynamics of the kicked soliton is the $4$-step Runge-Kutta
method.

The numerical simulations reveal that the soliton in the type I system can
be mobilized in the region where the total power $P$ satisfies $dW/dP<0$. In
this region, the soliton can be moved with a kick if $\eta >\eta _{c}$,
where $\eta _{c}$ increases as $P$ increases. A typical example for the
moving soliton in this case is depicted in Fig. \ref{moving}(a). When the
total power of the soliton is within the region of $dW/dP\sim 0$, $\eta _{c}$
becomes very large, and the soliton is destroyed under the strong kick. This
result implies that the solitons are immobile in this region. A typical
example of the soliton destroyed by the strong kick in this case is
illustrated in Fig. \ref{moving}(c). For the soliton in the type II system,
the numerical simulations show that the soliton can be mobilized before the
phase transition ($P<P_{\mathrm{cr}}$), whereas the soliton becomes immobile
and is pinned down after the phase transition ($P>P_{\mathrm{cr}}$). If the
strength of the kick is small, the soliton cannot be moved by the kick,
while if the strength of the kick is large, the kick will destroy the
soliton. Typical examples of moving and destroyed solitons are displayed in
Fig. \ref{moving}(b,d).

Based on the above description, the mobilities of the two types of solitons
can be clearly identified via the curves of $W(P)$ in Fig. \ref{character}%
(b). In Fig. \ref{character}(b), the solitons on the left side of the
vertical dotted line can be mobilized, while the solitons become immobile
(or destroyed) on the right side of the vertical dotted line. For
convenience, we define the location of the dotted line as $P=P_{\mathrm{d}}$%
. For the type I soliton, $P_{\mathrm{d}}=2.0$, while for the type II
soliton, $P_{\mathrm{d}}=P_{\mathrm{cr}}=3.24$. Consequently, the discrete
soliton in the current system can be mobilized when $P<P_{\mathrm{d}}$,
while the soliton cannot be moved by a kick; alternatively, the soliton can
be destroyed by exerting kicks of different strengths. Because moving
solitons exist on the left side of the vertical dotted line, it is implied
that solitons with a larger width can be mobilized more easily than narrower
width solitons, which follows from the traditional discrete system. However,
the border between mobile and immobile solitons is easier to distinguish in
the current system than\ in traditional discrete systems.

Haveing a kick of strength $\eta =\pi $, is equivalent to mapping the
solitons in the center of the Brillouin zone to the edge of the Brillouin
zone. Under this circumstance, two families of solitons with $P<P_{\mathrm{d}%
}$ and $P>P_{\mathrm{d}}$ exhibit different behaviors. When $P<P_{\mathrm{d}}
$, the two families of solitons are destroyed by implementing such a kick;
typical examples for this case are shown in Fig. \ref{kickpi}(a,b). When $%
P>P_{\mathrm{d}}$, the two families of solitons can remain localized and
propagate in a straight direction by exerting such a kick. This result
implies that a soliton solution may exist for both types of systems at the
edge of the Brillouin zone.

\subsection{Collision between moving solitons}

\begin{figure}[h]
{\includegraphics[width=1.0\columnwidth]{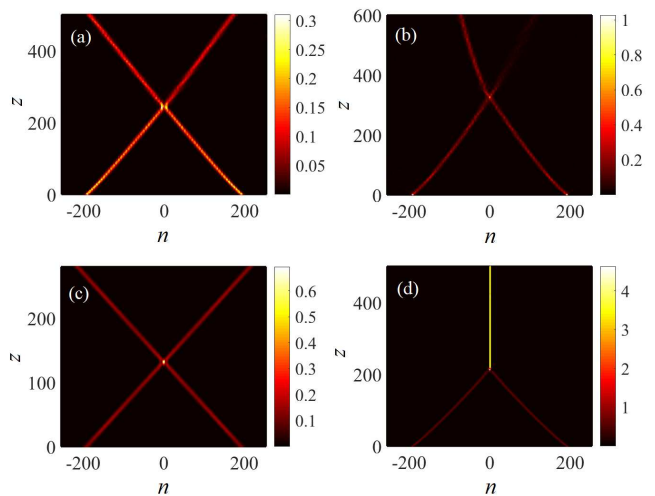}}
\caption{Typical examples of colliding solitons. (a) Elastic collision
between two type I solitons with $P_{\pm n_{0}}=1$, where $\protect\eta =0.1%
\protect\pi $. (b) Inelastic collision between two type I solitons with $%
P_{\pm n_{0}}=1.5$, where $\protect\eta =0.15\protect\pi $. (c) Elastic
collision between two type II solitons with $P_{\pm n_{0}}=1.5$, where $%
\protect\eta =0.1\protect\pi $. (d) Inelastic collision between two type II
solitons with $P_{\pm n_{0}}=3.2$, where $\protect\eta =0.1\protect\pi $.
Here, we select $n_{0}=192$ for all of the panels.}
\label{collision}
\end{figure}

Another nontrivial issue for the dynamics of the discrete soliton is
considering the collision between two moving solitons. Generally, the inital
states for both moving solitons can be constructed as
\begin{equation*}
u_{n,z=0}=\phi _{n+n_{0}}e^{i\eta (n+n_{0})}+\phi _{n-n_{0}}e^{-i\eta
(n-n_{0})},
\end{equation*}%
where $\phi _{n\pm n_{0}}$ denotes the stationary solutions centered at $%
n=\mp n_{0}$. The collisions between moving solitons are studied via the
soliton that can be mobilized (i.e., $P_{\pm n_{0}}<P_{\mathrm{d}}$).
Because the width of the moving soliton is relatively large, we need to
select a sufficiently large $n_{0}$ to prevent unnecessary overlap between
the two solitons. The numerical simulations show that the properties of the
colliding soliton pairs for both types of solitons are strongly influenced
by the relationship between $P_{\pm n_{0}}$ and $P_{\mathrm{d}}$. If $%
P_{+n_{0}}+P_{-n_{0}}<P_{\mathrm{d}}$, quasi-elastic collisions are observed
between soliton pairs. However, if $P_{+n_{0}}+P_{-n_{0}}>P_{\mathrm{d}}$,
inelastic collisions may be excited. This phenomenon can be explained as
follows: when the total power of the colliding solitons exceeds $P_{\mathrm{d%
}}$, the solitons may become immobile when they meet and merge together,
which results in an inelastic collision. Typical examples of these two
collisions for the two types of solitons are displayed in Fig. \ref%
{collision}.

\section{Conclusions}

The objective of this work is to study the characteristics and dynamics of
discrete solitons in zigzag waveguide arrays that feature different types of
linear mixing between nearest-neighbor and next-nearest-neighbor couplings.
Two families of discrete solitons are found in two types of waveguide
systems, where the nearest-neighbor coupling and the next-nearest-neighbor
coupling feature opposite relationships (i.e., $C_{1}\cdot C_{2}<0$). The
type I system is formed when the nearest-neighbor and next-nearest-neighbor
couplings feature negative and positive linear mixing, respectively, while
the type II system is formed when the nearest-neighbor and
next-nearest-neighbor couplings feature linear mixing opposite to those of
the type I system. The dispersion relations for these two types of settings
are analyzed by linearizing the system through the plane wave approximation,
and the characteristics of the soliton are studied by numerical simulation.%
\textbf{\ }The solitons in the type I system have a multipeak intensity
profile, and a $\pi $ phase shift exists for two adjacent peaks of a
soliton. The solitons in the type II system are staggered solitons.
Interestingly, a phase transition of the second kind occurs as the total
power of the soliton increases in the type II system. Furthermore, the phase
transition point, $P_{\mathrm{cr}}$, is clearly identified. Before the phase
transition, the intensity profile of the soliton in the type II system has a
Gaussian shape; after the phase transition, the majority of the power of the
soliton shrinks at the central point. Moreover, the mobilities of the two
families of solitons are studied throughout the paper. The solitons in the
current systems can be mobilized when their total power satisfies $P<P_{%
\mathrm{d}}$, and the solitons become immobile (pinned down or destroyed
under different kick strengths) when $P>P_{\mathrm{d}}$. The values of $P_{%
\mathrm{d}}$ for the two types of solitons are identified in this paper. For
the soliton in the type I system, $P_{\mathrm{d}}$ is the border between the
regions satisfying $dW/dP<0$ (where $W$ is the effective width of the
soliton) and $dW/dP\sim 0$. For the solitons in the type II system, $P_{%
\mathrm{d}}=P_{\mathrm{cr}}$. Collisions between moving solitons are also
discussed. An elastic collision is obtained if the total power of the
solitons is smaller than $P_{\mathrm{d}}$ when the solitons merge;
otherwise, an inelastic collision occurs.

Synthetic gauge fields and discrete matter-wave solitons are active research
topics and attract considerable interest in the field of Bose-Einstein
condensates \cite%
{Trombettoni2001,Santos2004,Atala2014,RW2014,Gligoric,Huaiyu2016,Zhiwei2017,Yongyao2017,Qingzhou2018,PRL_LYY}%
. The discussion in this paper is also suitable for a matter-wave soliton
trapped in a zigzag optical lattice with the same competition between the
two types of coupling (i.e., the hopping rate).

\section{Acknowledgments}

This work was supported by the NNSFC (China) through grant Nos. 11905032,
11874112, and 11575063, the Foundation for Distinguished Young Talents in
Higher Education of Guangdong through grant No. 2018KQNCX279, Key Research Projects of General Colleges in Guangdong Province through grant No. 2019KZDXM001.

\textbf{Declaration of competing interest}

The authors declare that they have no known competing financial interests or
personal relationships that could have appeared to influence the work
reported in this paper.

\end{document}